\documentclass[aps,prd,preprint,preprintnumbers,unsortedaddress,superscriptaddress,showpacs,nofootinbib]{revtex4-1}

\usepackage{graphicx,color}
\usepackage{relsize}
\usepackage{slashed}
\usepackage{color}
\usepackage{ulem}
\usepackage{tabu}
\usepackage{bm}

\usepackage{amsmath}
\usepackage{amssymb}
\usepackage{amsthm}
\usepackage{mathrsfs}
\usepackage{graphicx}
\usepackage{epstopdf}
\usepackage{fancyhdr}
\usepackage{array}
\usepackage[all]{xy}
\usepackage{eufrak}
\usepackage{euscript}
\usepackage{enumerate}
\usepackage{slashed}
\usepackage{hyperref}
\usepackage{subfigure} 
\usepackage{epstopdf} 

\hypersetup{pdftex,colorlinks=true,linkcolor=blue,citecolor=blue,menucolor=black,urlcolor=blue,filecolor=blue}

\newcommand{\myvec}[1]{\ensuremath{\begin{pmatrix}#1\end{pmatrix}}}

\begin{document}

\title{Molecular $ \boldsymbol{\Omega_b }$ states}

\author{Wei-Hong Liang}
\email{liangwh@gxnu.edu.cn}
\affiliation{Department of Physics, Guangxi Normal University, Guilin 541004, China}

\author{J.~M.~Dias}
\email{jdias@if.usp.br}
\affiliation{Departamento de F\'{i}sica Te\'{o}rica and IFIC, Centro Mixto Universidad de Valencia - CSIC,
Institutos de Investigaci\'{o}n de Paterna, Aptdo. 22085, 46071 Valencia, Spain.}
\affiliation{Instituto de F\'{i}sica, Universidade de S\~{a}o Paulo, Rua do Mat\~{a}o, 1371, Butant\~{a}, CEP 05508-090, S\~{a}o Paulo, S\~{a}o Paulo, Brazil}

\author{V.~R.~Debastiani}
\email{vinicius.rodrigues@ific.uv.es}
\affiliation{Departamento de F\'{i}sica Te\'{o}rica and IFIC, Centro Mixto Universidad de Valencia - CSIC,
Institutos de Investigaci\'{o}n de Paterna, Aptdo. 22085, 46071 Valencia, Spain.}

\author{E.~Oset}
\email{eulogio.oset@ific.uv.es}
\affiliation{Departamento de F\'{i}sica Te\'{o}rica and IFIC, Centro Mixto Universidad de Valencia - CSIC,
Institutos de Investigaci\'{o}n de Paterna, Aptdo. 22085, 46071 Valencia, Spain.}

\date{\today}

\begin{abstract}
  Motivated by the recent finding of five $\Omega_c$ states by the LHCb collaboration, and the successful reproduction of three of them in a recent approach searching for molecular states of meson-baryon with the quantum numbers of $\Omega_c$, we extend these ideas and make predictions for the interaction of meson-baryon in the beauty sector, searching for poles in the scattering matrix that correspond to physical states. We find several $\Omega_b$ states: two states with masses 6405~MeV and 6465~MeV for $J^P= \frac{1}{2}^-$; two more states with masses 6427~MeV and 6665~MeV for $\frac{3}{2}^-$; and three states between 6500 and 6820~MeV, degenerate with $J^P=\frac{1}{2}^-,\,\frac{3}{2}^-$, stemming from the interaction of vector-baryon in the beauty sector.
\end{abstract}



\maketitle

\section{Introduction}
\label{sec:intro}

The study of baryon states with charm or beauty is capturing much attention in hadron physics recently \cite{slz,chenrep,stonepen,zhoupen,gengrev,olsenrev}. The finding of baryon states of hidden charm (pentaquarks) in Refs.~\cite{lhcpen,lhcpen2} certainly stimulated this field, but this was followed by another relevant discovery, with the observation of five new $\Omega_c$ states \cite{omegac}. This discovery stimulated much theoretical work trying to describe these states (see detailed description of works in Ref.~\cite{omeus}). From these we specially refer to Ref.~\cite{montana} which uses a dynamics closely related to ours.

The $\Omega_b$ states have not been the subject of much study. Experimentally the PDG \cite{pdg2016} quotes the ground state $J^P=\frac{1}{2}^+$ and no more states. The $J^P=\frac{3}{2}^+$ excited state has not yet been observed. Theoretical work is also scarce, but there are predictions for $\frac{1}{2}^-, \frac{3}{2}^-$, $\frac{5}{2}^-$, $\frac{1}{2}^+$, $\frac{3}{2}^+$, $\frac{5}{2}^+$, $\frac{7}{2}^+$ states in different quark models, relativistic quark model \cite{rqm}, non-relativistic quark model \cite{nrqm} and QCD sum rules \cite{hosaka1,hosaka2,hosaka3,Albuquerque:2009pr}.
In this latter line, some recent works make predictions for the
$\Omega_b$ ground state and the first orbitally and radially excited
states, with spin $J=1/2$ and $3/2$ \cite{azizi1,azizi2}.

Unlike in the charm sector, no work on $\Omega_b$ molecular states has been done. The recent finding of the $\Omega_c$ states by the LHCb collaboration \cite{omegac}, and the steady work in the search of new states, makes the study of $\Omega_b$ molecular states relevant and opportune. The opportunity is even more apparent after the realization in the works of Refs.~\cite{montana,omeus} that several of the observed states can be well described in the molecular picture. In Ref.~\cite{montana} the local hidden gauge formalism is used to obtain the interaction between mesons and baryons in the charm sector, with the quantum number of $\Omega_c$. The channels considered are $ \Xi_c \bar K, \,  \Xi'_c \bar K, \, \Xi D , \,  \Omega_c \eta, \,  \Omega_c \eta'$ and the interaction proceeds via the exchange of vector mesons, extending the local hidden gauge Lagrangian \cite{hidden1,hidden2,hidden4} to SU(4). Two states of $\frac{1}{2}^-$ could be associated to the $\Omega_c(3050)$ and $\Omega_c(3090)$ states of Ref.~\cite{omegac}, both in energy and approximately width. The work of Ref.~\cite{omeus} continues with this line but does not assume SU(4) for the interaction. Instead, the wave functions of the charmed baryons in the coupled channels isolate the charm quark and symmetrize the spin-flavor wave function of the light quarks. The diagonal terms in the transition potentials between the coupled channels coincide in Ref.~\cite{montana} and Ref.~\cite{omeus}, but there are differences in the non-diagonal ones. In addition in Ref.~\cite{omeus} the states $\Xi^*_c \bar K, \,  \Omega^*_c \eta, \,  \Xi^* D$ are considered from where another state of $\frac{3}{2}^-$ emerges that can also be associated to a third state of Ref.~\cite{omegac}.

In Ref.~\cite{omeus} one obtains three states, which can be associated to three states of Ref.~\cite{omegac}, and the agreement of masses and widths is good. The success of this approach to get some of the observed $\Omega_c$ states stimulates us to use the same idea and make predictions for $\Omega_b$ states. The work is simple because all one must do is to change a $c$ quark by a $b$ quark, and the matrix elements of the interaction are formally the same, although some differences appear in the non-diagonal terms due to the different masses of the hadrons. On the other hand, we can also benefit from the works of Refs.~\cite{ozpineci,geng} which show that to preserve heavy quark symmetry in the molecular states one should use a common cut-off in the regularization of the meson-baryon loop functions in the heavy quark sector. An alterative method to preserve this symmetry is provided in Ref.~\cite{alten}. Given these two ingredients, we feel confident that the two $\frac{1}{2}^-$ states and the $\frac{3}{2}^-$ state that come from our approach for the $\Omega_b$ should be realistic. We also predict other states at higher masses, analogous to some states predicted for $\Omega_c$ which lie in a region of large background and are more difficult to identify. $\Omega_b$ states are produced with smaller statistics in LHCb but they are subject of investigation. With increased luminosity in the next LHCb runs, the observation of $\Omega_b$ states will become a state of the art and the comparison with the predictions done here will shed light on hadron dynamics and the nature of some hadronic states.

\section{Formalism}
\label{sec:form}

We follow closely the formalism of Ref.~\cite{omeus} by changing a $c$ quark by a $b$ quark. For the case of $\Omega_c$ we took the coupled channels from Ref.~\cite{romanets} up to an energy of 3470 MeV, far above the energy of the states seen in Ref.~\cite{omegac}. In the present case we take the corresponding states changing the quark $c$ by the $b$ quark. We take into account the $S$-wave interaction of these coupled channels and hence we can have states with $J^P= \frac{1}{2}^-, \frac{3}{2}^-$. In Tables \ref{tab:tab1}, \ref{tab:tab2} and \ref{tab:tab3} we show these coupled channels and the corresponding threshold masses \footnote{The $\Omega^*_b$ state has not yet been observed. We estimate its mass as follows. In the charm sector, we have $m_{D^*}-m_D = 142\, {\rm MeV}$, $m_{\Omega^*_c}-m_{\Omega_c} = 71\, {\rm MeV}$. Hence the difference of masses between $\Omega^*_c$ and $\Omega_c$ is about one half the one between $D^*$ and $D$. We apply the same rule in the $b$ sector and take $m_{\Omega^*_b} - m_{\Omega_b} \simeq \frac{1}{2} (m_{B^*}-m_B) \simeq 23\, {\rm MeV}$. If we assume, following the rules of heavy quark spin symmetry, that $m_{\Omega^*_b} - m_{\Omega_b}$ goes as $\frac{1}{m_b}$ and $m_{\Omega^*_c} - m_{\Omega_c} \sim \frac{1}{m_c}$, then we get  $m_{\Omega^*_b} - m_{\Omega_b} \simeq 28\, {\rm MeV}$. We take the average value $25\, {\rm MeV}$, hence $m_{\Omega^*_b} = 6071\, {\rm MeV}$.}.

\vspace{0.6cm}
\begin{table}[h]
\caption{The pseudoscalar-baryon states with $J^P=\frac{1}{2}^-$ and their threshold masses in MeV.}
\centering
\begin{tabular}{c | c c c c}
\hline\hline
{\bf States} ~& ~~~$\Xi_b\bar{K}$~~ & ~~$\Xi^{\prime}_b\bar{K}$~~ & ~~$\Omega_b \eta$~~ & ~~$\Xi \bar B$~~\\
\hline
{\bf Threshold} ~& $6289$ & $6431$ & $6594$ & $6598$ \\
\hline\hline
\end{tabular}
\label{tab:tab1}
\end{table}
\begin{table}[h!]
\caption{The pseudoscalar-baryon states with $J^P=\frac{3}{2}^-$ and their threshold masses in MeV.}
\centering
\begin{tabular}{c | c c c}
\hline\hline
{\bf States} ~& ~~~$\Xi^*_b \bar{K}$~~ & ~~$\Omega^*_b\eta$~~ & ~~$\Xi^* \bar B$~~ \\
\hline
{\bf Threshold} ~& $6451$ & $6619$ & $6813$ \\
\hline\hline
\end{tabular}
\label{tab:tab2}
\end{table}
\begin{table}[h!]
\caption{The vector-baryon states with $J^P=\frac{1}{2}^-,\, \frac{3}{2}^-$ and their threshold masses in MeV.}
\centering
\begin{tabular}{c | c c c}
\hline\hline
{\bf States} ~& ~~~$\Xi \bar{B}^*$~~ & ~~$\Xi_b \bar{K}^*$~~ & ~~$\Xi'_b \bar K^*$~~\\
\hline
{\bf Threshold} ~& $6643$ & $6687$ & $6829$ \\
\hline\hline
\end{tabular}
\label{tab:tab3}
\end{table}

\begin{figure}[b!]
\begin{center}
\includegraphics[scale=0.7]{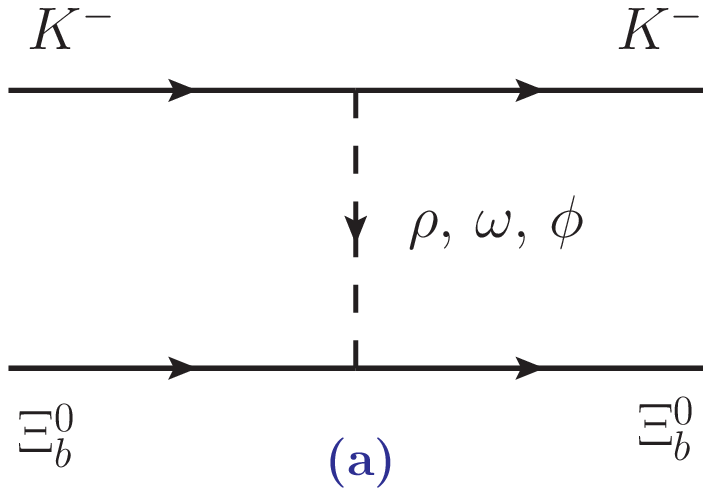}
\hspace{0.9cm}
\includegraphics[scale=0.7]{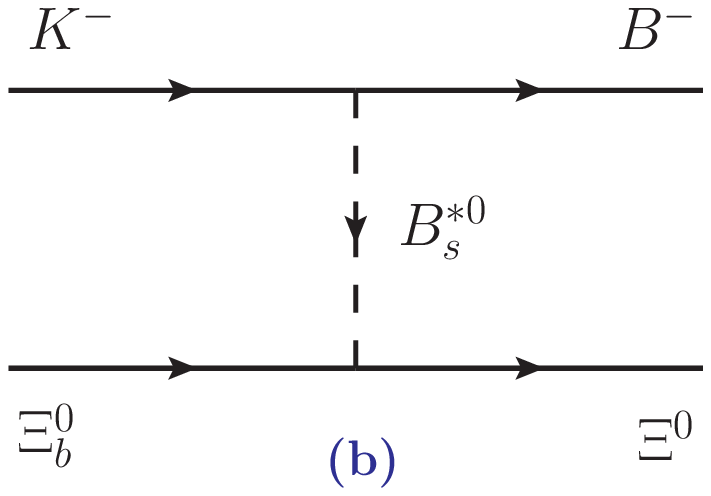}
\end{center}
\vspace{-0.7cm}
\caption{Vector mesons exchanged in the diagonal transition of $K^- \Xi^0_b \to K^- \Xi^0_b$ (a) and non-diagonal one of $K^- \Xi^0_b \to B^- \Xi^0$ (b).}
\label{Fig:1}
\end{figure}

The interaction between these channels at tree level is obtained using the local hidden gauge (LHG) approach  \cite{hidden1,hidden2,hidden4} extended to the charm and beauty sector. The interaction is mediated by the exchange of vector mesons, as shown in Fig. \ref{Fig:1} for two cases.

The uper vertex in Fig. \ref{Fig:1} for pseudoscalar(P)-pseudoscalar(P)-vector(V) is given in terms of the Lagrangian
\begin{equation}\label{eq:Lagrangian}
  \mathcal{L}_{\rm VPP}=-ig \langle \, [\Phi\, , \partial_{\mu}\Phi ] \, V^\mu \, \rangle,
\end{equation}
with
$g$ the coupling of the LHG approach,
$g= \frac{M_V}{2f_\pi} $ ($m_V \sim 800 \,{\rm MeV}$, $f_\pi=93 \,{\rm MeV}$),
and $\Phi$, $V$ the SU(4) matrices for pseudoscalar mesons and the vector mesons with the quarks $u, d, s, b$. The matrices are given explicitly in Ref.~\cite{omeus} for the $u, d, s, c$ sector, and the analogues for the $u, d, s, b$ quarks are easily constructed. There is no need to write them. The change of the $c$ quark by the $b$ quark, with the same structure of the wave functions,  has as a consequence that the matrix elements of the transitions are formally the same up to some small change in the non-diagonal terms that we discuss below.

It was shown in Ref.~\cite{rocasakai} that in the case of the diagonal terms one exchanges light vectors and the heavy quark ($b$ quark here) is a spectator. This guarantees that we can obtain these terms from the SU(3) sector of the LHG approach \cite{hidden1,hidden2,hidden4} and that the interaction does not depend on the heavy quark, then automatically satisfying the heavy quark symmetry \cite{neubert,manohar}. In the non-diagonal transitions, sometimes a $B^*_s$ vector is exchanged, instead of $\rho, \omega, \phi$ in the diagonal terms. Then these terms are very much suppressed by the mass of the $B^*_s$. In Ref.~\cite{omeus}, a $D^*_s$ was exchanged, and both in Ref.~\cite{montana} and Ref.~\cite{omeus} it was found that this was penalized by a factor about $1/4$. We follow the same argumentation as in Ref.~\cite{omeus} and write for the heavy propagator compared with the diagonal case
\begin{eqnarray}
\lambda = \frac{-M_V^2}{{q^{0}}^2-|\vec{q}\,|^2 -m^2_{B^*_s}} \simeq \frac{-M^2_{V}}{(m_K - m_B)^2 - m^2_{B^*_s}}\approx 0.1 \, .
\end{eqnarray}

The lower vertices in Fig.~\ref{Fig:1} for the vector(V)-baryon(B)-baryon(B) couplings are obtained by writing explicitly the vector mesons and the baryons in terms of quarks, as it is done in Ref.~\cite{omeus}. The wave functions for the baryons are identical to those used in Ref.~\cite{omeus} changing the $c$ quark by a $b$ quark. As to the isospin states (we need only $I=0$ to construct $\Omega_b$ state), we have the isospin multiplets
%
\begin{alignat*}{4}
& \bar{K} = \myvec{\bar{K}^0\\-K^-};  & \qquad   & \bar B = \myvec{\bar B^0\\-B^-};  & \qquad  & \Xi = \myvec{\Xi^0\\-\Xi^-};  & \qquad
&  \Xi^* = \myvec{\Xi^{*0} \\\Xi^{*-}}; \\ 
\vspace{3cm}
& \Xi_b = \myvec{\;\Xi_b^0\\ \;\Xi_b^-};  & \qquad   &   \Xi'_b = \myvec{\;\Xi'^{\,0}_b\\ \; \Xi'^{\,-}_b};  & \qquad   & \Xi^{*}_b = \myvec{\Xi^{* 0}_b\\ \Xi^{*-}_b} ; & \qquad   &
\end{alignat*}
and the isospin $I=0$ states are easily obtained from those and have the same relative signs as in the charm sector in Ref.~\cite{omeus}.

Then the transition potentials are given by
\begin{eqnarray}
\label{eq:kernel}
V_{ij}= D_{ij} \,  \frac{1}{4f^2_\pi}\,(p^0+p'^0),
\end{eqnarray}
where $p^0, p'^0$ are the energies of the incoming and outgoing mesons, and the $D_{ij}$ coefficients are given in Tables \ref{tab:DijPB}, \ref{tab:DijPB2} and \ref{tab:DijVB}.
\begin{table}[h!]
\caption{$D_{ij}$ coefficients for the PB channels with $J^P=\frac{1}{2}^-$.}
\centering
\begin{tabular}{c || c c c c}
\hline\hline
 $J^P=1/2^-$~ & ~~$\Xi_b\bar{K}$~ & ~$\Xi^{\prime}_b\bar{K}$~ & ~~$\Xi \bar B$~ & $\Omega_b \eta$ \\
\hline\hline
$\Xi_b\bar{K}$ & $-1$ & $0$ & $-\frac{1}{\sqrt{2}}\lambda$ &~$0$\\
$\Xi^{\prime}_b\bar{K}$ &  & $-1$ & ~~$\frac{1}{\sqrt{6}}\lambda$ & $-\frac{4}{\sqrt{3}}$ \\
$\Xi \bar B$ & & & $-2$ & ~$\frac{\sqrt{2}}{3}\lambda$~~ \\
$\Omega_b \eta$ & & &  & $0$ \\
\hline\hline
\end{tabular}
\label{tab:DijPB}
\end{table}
\begin{table}[h!]
\caption{$D_{ij}$ coefficients for the PB channels with $J^P=\frac{3}{2}^-$.}
\centering
\begin{tabular}{c || c c c}
\hline\hline
 $J^P=3/2^-$~ & ~~$\Xi^*_b \bar{K}$~ & ~$\Omega^*_b \eta$~ & ~$\Xi^* \bar B$ \\
\hline\hline
$\Xi^*_b \bar{K}$ & $-1$ & $-\frac{4}{\sqrt{3}}$ & $\frac{2}{\sqrt{6}}\lambda$\\
$\Omega^*_b \eta$ &  & $0$ & $-\frac{\sqrt{2}}{3}\lambda$  \\
$\Xi^* \bar B$ & & & $-2$  \\
\hline\hline
\end{tabular}
\label{tab:DijPB2}
\end{table}
\begin{table}[h!]
\caption{$D_{ij}$ coefficients for the VB channels with $J^P=\frac{1}{2}^-, \, \frac{3}{2}^-$.}
\centering
\begin{tabular}{c || c c c}
\hline\hline
 $J^P=1/2^-, \, 3/2^-$~ & ~~$\Xi \bar{B}^*$~ & ~$\Xi_b \bar{K}^*$~ & ~$\Xi'_b \bar K^*$ \\
\hline\hline
$\Xi \bar{B}^*$ & $-2$ & $-\frac{1}{\sqrt{2}}\lambda$ & $\frac{1}{\sqrt{6}}\lambda$\\
$\Xi_b \bar{K}^*$ &  & $-1$ & $0$  \\
$\Xi'_b \bar K^*$ & & & $-1$  \\
\hline\hline
\end{tabular}
\label{tab:DijVB}
\end{table}
By including relativistic correction in $S$-wave \cite{Bennhold}, the transition potentials $V_{ij}$ can be written as
\begin{eqnarray}
\label{eq:kernel2}
V_{ij}= D_{ij} \,  \frac{2 \sqrt{s}-M_{B_i}-M_{B_j}}{4f^2_\pi}\, \sqrt{\frac{M_{B_i}+E_{B_i}}{2 M_{B_i}}}
\, \sqrt{\frac{M_{B_j}+E_{B_j}}{2 M_{B_j}}} ,
\end{eqnarray}
with $M_{B_i}$ and $E_{B_i}$ the mass and the center-of-mass energy of the baryon in the $i$-th channel.

From the potential of Eq. \eqref{eq:kernel2} we construct the $t$-matrix using the Bethe-Salpeter equation in coupled channels in the on-shell factorized form \cite{nsd,ollerulf}
\begin{equation}\label{eq:BSEQ}
  T= [1-VG]^{-1}V,
\end{equation}
where $G$ is the diagonal matrix from the loop function of the intermediate meson-baryon propagators. As in
Ref.~\cite{omeus} we take the cut-off regularization and use here the same cut-off that was used in Ref.~\cite{omeus} to respect rules of heavy quark symmetry, as discussed in Refs.~\cite{ozpineci,geng}. We take $q_{max}=650 \, {\rm MeV}$, which was the cut-off providing good agreement with the experimental states in Ref.~\cite{omeus}. The poles of the amplitudes provide the states and they are searched in the second Riemann sheet of the complex energy plane as in Ref.~\cite{omeus}.

It should be noted that in the case of vector-baryon interaction,
$V_{ij}$ of Eq.~\eqref{eq:kernel2} has the extra factor $\vec \epsilon \cdot \vec \epsilon\, '$,
where $\vec \epsilon$ and $\vec \epsilon\, '$ are
the polarization vectors of the initial and final vector mesons, stemming from the
V($\vec \epsilon$)-V($\vec \epsilon\, '$)-V(virtual) vertex
in the limit of small three momenta compared to the vector meson masses \cite{angel,omeus}.
This factor induces degeneracy in $J^P= \frac{1}{2}^-,\, \frac{3}{2}^-$ vector-baryon($\frac{1}{2}^+$)
states in $S$-wave.
This factor does not appear for pseudoscalar mesons and, hence,
for pseudoscalar-baryon($\frac{1}{2}^+$) we only obtain $J^P= \frac{1}{2}^-$ states
and for the pseudoscalar-baryon($\frac{3}{2}^+$) we only obtain $J^P= \frac{3}{2}^-$ states.
For the case of vector-baryon we predict states both in $\frac{1}{2}^-$ and $\frac{3}{2}^-$,
which in our approach have the same energy.
The degeneracy can be broken by mixing the pseudoscalar-baryon and vector-baryon channels,
which is done through pion exchange \cite{Garzon},
but in our case pion exchange is found to be small.

\section{Results}
\label{sec:res}

In Tables \ref{tab:resPB1}, \ref{tab:resPB2} and \ref{tab:resVB}, we show the results.
In Table \ref{tab:resPB1} we see that we obtain two states with $J^P=\frac{1}{2}^-$ at 6405 MeV and 6465 MeV. The widths are given by twice the imaginary part of the pole position, and they are small in all cases.
We also show the couplings of the states obtained to the different coupled channels, as well as the product $g_i \, G_i^{II}$ ($G_i^{II}$ is the $G$ function calculated at the pole in the second Riemann sheet), which as shown in Ref.~\cite{gamejuan} is proportional to the wave function at the origin. By looking at the couplings and the wave function at the origin we can see that the first state, at 6405 MeV, couples strongly to $\Xi'_b \bar K$ and next to $\Omega_b \eta$. The second state, at 6465 MeV, couples most strongly to $\Xi \bar B$ and little to the other channels, and hence it qualifies as mostly a $\Xi \bar B$ bound state.

\begin{table}[tbh]
\caption{The poles, and coupling constants of the poles to various channels in the PB sector with $J^P={1/2}^-$, taking $q_{max}=650$ MeV. $g_i$ has no dimension and $g_i\,G^{II}_i$ has dimension of MeV.}
\centering
\begin{tabular}{c c c c c}
\hline\hline
$\boldsymbol{6405.2}$~ & $\Xi_b\bar{K} $ & $\Xi'_b\bar{K} $ & $\Xi \bar B $ & $\Omega_b \eta $ \\
\hline
$g_i$ & ~$-0.01+i0.02$~ & ~~$\boldsymbol{2.04+i0.01}$~ & ~$-1.62+i0.02$~  & ~$\boldsymbol{2.08+i0.01}$~\\
$g_i\,G^{II}_i$ & $-0.34-i0.47$ & $\boldsymbol{-37.31-i0.18}$ & ~$2.27-i0.02$ & ~~$\boldsymbol{-18.28-i0.09}$ \\
\hline\hline
$\boldsymbol{6465.3+i1.2}$~ & $\Xi_b\bar{K}$ & $\Xi'_b\bar{K}$ & $\Xi \bar B$  &  $\Omega_b \eta$\\
\hline
$g_i$ & ~$0.07-i0.15$~ & ~~~~$0.11+i0.125$~ & ~~~$\boldsymbol{10.70-i0.10}$~ & ~$0.15+i0.11$~ \\
$g_i\,G^{II}_i$ & $3.92+i3.91$ & $-4.53-i1.66$ & $\boldsymbol{-18.89+i0.08}$  & ~~$-1.55-i1.14$\\
\hline\hline
\end{tabular}
\label{tab:resPB1}
\end{table}

\begin{table}[tbh]
\caption{The poles, and coupling constants of the poles to various channels in the PB sector with $J^P={3/2}^-$, taking $q_{max}=650$ MeV. $g_i$ has no dimension and $g_i\,G^{II}_i$ has dimension of MeV.}
\centering
\begin{tabular}{c c c c}
\hline\hline
$\boldsymbol{6427.1}$~ & ~$\Xi^*_b \bar{K} $ ~& ~$\Omega^*_b\eta $~ & ~$\Xi^* \bar B $~ \\
\hline
$g_i$ & ~$\boldsymbol{2.01}$~ & ~$\boldsymbol{2.05}$~ & ~$-0.60$~ \\
$g_i\,G^{II}_i$ & $\boldsymbol{-37.17}$ & ~~$\boldsymbol{-17.86}$ & ~~$0.53$ \\
\hline\hline
$\boldsymbol{6664.8+i0.2}$~ & $\Xi^*_b \bar{K}$ & $\Omega^*_b\eta$ & $\Xi^* \bar B$  \\
\hline
$g_i$ & ~$-0.02-i0.01$~ & ~$0.10+i0.05$~ & ~~~~$\boldsymbol{11.06+i0.01}$~  \\
$g_i\,G^{II}_i$ & ~~$0.59-i0.53$ & ~~$-3.07+i0.41$ & $\boldsymbol{-19.31-i0.02}$ \\
\hline\hline
\end{tabular}
\label{tab:resPB2}
\end{table}

\begin{table}[tbh]
\caption{The poles, and coupling constants of the poles to various channels in the VB sector with $J^P={1/2}^-,\, {3/2}^-$, taking $q_{max}=650$ MeV. $g_i$ has no dimension and $g_i\,G^{II}_i$ has dimension of MeV.}
\centering
\begin{tabular}{c c c c}
\hline\hline
$\boldsymbol{6508.0}$~ & $\Xi \bar{B}^* $ & $\Xi_b\bar{K}^* $ & $\Xi'_b \bar K^* $ \\
\hline
$g_i$ & ~~~$\boldsymbol{10.88}$~ & ~~~$0.32$~ & ~$-0.15$~ \\
$g_i\,G^{II}_i$ & $\boldsymbol{-18.86}$ & $-2.37$ & ~~$0.77$ \\
\hline
\hline
$\boldsymbol{6676.1+i0.1}$~ & $\Xi \bar{B}^*$ & $\Xi_b\bar{K}^*$ & $\Xi'_b \bar K^*$  \\
\hline
$g_i$ & ~$-0.05-i0.09$~ & ~~~$\boldsymbol{1.78-i0.10}$~ & ~~~$0.01+i0.01$~  \\
$g_i\,G^{II}_i$ & ~~$0.68+i0.27$ & $\boldsymbol{-35.16+i1.90}$ & $-0.07-i0.01$ \\
\hline\hline
$\boldsymbol{6817.5}$~ & $\Xi \bar{B}^*$ & $\Xi_b\bar{K}^*$ & $\Xi'_b \bar K^*$  \\
\hline
$g_i$ & ~$-0.01+i0.02$~ & ~$0.01-i0.01$~ & ~$\boldsymbol{1.77+i0.01}$~  \\
$g_i\,G^{II}_i$ & $-0.26-i0.01$ & $0.05+i0.03$ & $\boldsymbol{-34.71-i0.18}$ \\
\hline
\end{tabular}
\label{tab:resVB}
\end{table}

In Table \ref{tab:resPB2}, we show the states that appear exclusively in $J^P=\frac{3}{2}^-$. They are obtained from a pseudoscalar meson ($\eta, \,\bar K$ or $\bar B$) and a baryon of spin-parity $\frac{3}{2}^+$ ($\Xi^*_b, \, \Omega^*_b$ or $\Xi^*$).
We also find two states, one at 6427 MeV and the other one at 6665 MeV. The first one couples mostly to $\Xi^*_b \bar K$ but also to $\Omega^*_b \eta$, while the second one couples mostly to $\Xi^* \bar B$ and little to the other channels and would qualify as mostly a $\Xi^* \bar B$ bound state. We can see here a qualitative difference with the results obtained for the $\Omega_c$ states. There the first $\frac{3}{2}^-$ state had bigger energy than the two $\frac{1}{2}^-$ states. Here it is in the middle. The reason is simple because the difference between $\Xi^*_b$ and $\Xi_b$, $\Omega^*_b$ and $\Omega_b$ are now smaller than between $\Xi^*_c$ and $\Xi_c$, $\Omega^*_c$ and $\Omega_c$. Confirmation of this feature by future experiments would already provide support to the molecular picture that we discuss here.

Finally we also show in Table \ref{tab:resVB} the three states that we obtain from the vector-baryon channels. Here the spin-parity can be $\frac{1}{2}^-, \, \frac{3}{2}^-$, since in our approach these spin states are degenerate. One can break the degeneracy allowing pion exchange between the pseudoscalar-baryon and vector-baryon states as done in Refs.~\cite{xiao,uchino}, but as shown in Ref.~\cite{omeus} for the $\Omega_c$ states, the pion exchange was very small in this case. We obtain three states at 6508 MeV, 6676 MeV and 6818 MeV with very small width. The first state couples mostly to $\Xi \bar B^*$, the second one to $\Xi_b \bar K^*$ and the third one to $\Xi'_b \bar K^*$ with small coupling to the other channels, which make them qualify as bound states of these channels. The last state would correspond to a bound $\Xi'_b \bar K^*$ state,  with a binding of barely 11 MeV.

In Ref.~\cite{omeus} we had also found three states of vector-baryon nature, but they appear in a region where the experimental background is large. In any case our approach does not tell the strength at which the states are produced in the particular LHCb reaction. Yet, other more selective methods to produce the states like $\Omega_b \to (\Xi^+_c K^-) \pi^-$ suggested in Ref.~\cite{belyaev}, in the case of $\Omega_c$ states, would allow theoretical approaches to predict the relative strength at which every state is produced. In the present case we cannot rely on the weak decay of heavier resonances, but it would be convenient to search for the $\Omega_b$ states in different reactions where the strength for the production of the different states would be different, and one would have more chances to find them. The analogy of the $\Omega_b$ states with the $\Omega_c$ ones, suggest that three of the states predicted by us could be easily seen using a similar reaction.

\section{Conclusions}
\label{sec:conc}

Motivated by the recent experimental finding of five $\Omega_c$ states and the successful reproduction of three of these states in the molecular model for $\Omega_c$ states, we have used the same formalism used to obtain the $\Omega_c$ states to make predictions for $\Omega_b$ states, just changing one $c$ quark by a $b$ quark. The only freedom in the approach is the regulator in the loop function of the meson-baryon states, but for this we used a cut-off regularization with the same cut-off as in the charm sector, which has been shown in different approaches to respect the rules of heavy quark symmetry. In this way, the predictions of the model should be very accurate. Since we study only the interaction of meson-baryon in $S$-wave, we predict states of pseudoscalar-baryon nature with $J^P=\frac{1}{2}^-$ and $\frac{3}{2}^-$, and we find two states of $\frac{1}{2}^-$, two states of $\frac{3}{2}^-$ and three more states, degenerate in our approach in  $\frac{1}{2}^-, \, \frac{3}{2}^-$, that stem from the interaction of vector mesons with baryons.

Although the production of $\Omega_b$ states is done with less statistics than the $\Omega_c$ states in LHCb, with increased luminosity in future runs the access to these states will become a state of the art. Predictions done before the experiment are very useful, and comparison of the results obtained here with experimental measurements in the future will help us understand better hadron dynamics and the nature of some of the states found.

\vspace{-0.3cm}
\begin{acknowledgments}
V. R. Debastiani acknowledges the Programa Santiago Grisolia of Generalitat
Valenciana (Exp. GRISOLIA/2015/005). J. M. Dias thanks the Funda\c c\~ao de
Amparo \`a Pesquisa do Estado de S\~ao Paulo (FAPESP) for support by FAPESP
grant 2016/22561-2.
This work is partly supported by the National Natural Science Foundation of China (NSFC) under Grant Nos. 11565007, 11747307 and 11647309. This work is also partly supported by the Spanish Ministerio de Economia y Competitividad
and European FEDER funds under the contract number FIS2011-28853-C02-01, FIS2011-28853-C02-02, FIS2014-57026-REDT, FIS2014-51948-C2-1-P, and FIS2014-51948-C2-2-P, and the Generalitat Valenciana in the program Prometeo II-2014/068.
\end{acknowledgments}


  \end{document}